\documentclass[12pt,preprint]{aastex}
\usepackage[utf8]{inputenc} 
 \usepackage[russian,english]{babel} 
\usepackage{float}

\begin{document}

\title{Experimental study of an advanced concept of moderate-resolution holographic spectrographs}

\author{Eduard Muslimov\altaffilmark{1,2},
Gennady Valyavin\altaffilmark{3},
Sergei Fabrika\altaffilmark{3,4},
Faig Musaev\altaffilmark{3},
Gazinur Galazutdinov\altaffilmark{5,3,6},
Nadezhda Pavlycheva\altaffilmark{2},
Eduard Emelyanov\altaffilmark{3}
}

\altaffiltext{1}{Aix Marseille Univ, CNRS, LAM, Laboratoire d'Astrophysique de Marseille, Marseille, France, 38 rue Joliot-Curie, Marseille 13388, France}
\altaffiltext{2}{Kazan National Research Technical University named after A.N. Tupolev--KAI, 10 K. Marx, Kazan 420111, Russia}
\altaffiltext{3}{Special Astrophysical Observatory, Russian Academy of Sciences, Nizhnii Arkhyz 369167, Russia}
\altaffiltext{4}{Kazan Federal University, Kremlevskaya 18, Kazan 420008, Russia}
\altaffiltext{5}{Instituto de Astronomia, Universidad Catolica del Norte, Av. Angamos 0610, Antofagasta 1270709, Chile}
\altaffiltext{6}{Central (Pulkovo) Observatory, Pulkovskoe Shosse 65, Saint-Petersburg, Russia 196140}

\begin{abstract}
We present the results of an experimental study of an advanced moderate-resolution spectrograph based on a cascade of narrow-band
holographic gratings. The main goal of the project is to achieve a moderately high spectral resolution with R up to ~5000 simultaneously in the
4300-6800\,\AA\, visible spectral range on a single standard CCD, together with an increased throughput. The experimental study consisted of (1)
resolution and image quality tests performed using the solar spectrum; and (2) a total throughput test performed for a number of wavelengths using a
calibrated lab monochromator. The measured spectral resolving power reaches values over R$>$4000 while the experimental throughput is as
high as 55\%, which is in good agreement with the modeling results. Comparing the obtained characteristics of the spectrograph under consideration
with the best existing spectrographs, we conclude that the used concept can be considered a very competitive and cheap alternative to the existing
spectrographs of the given class. We propose several astrophysical applications for the instrument and discuss the prospect of creating its
full-scale version.
\end{abstract}

\section{Introduction}
\label{Intro}
The optical design of any diffraction spectrograph can rely on one of two basic approaches \citep{Palm}. The first approach relies on the use of a grating working in a
single (usually the +1$^{st}$) diffraction order, whereas the second one relies on echelle gratings working in
multiple diffraction orders. In the first case, both the diffraction efficiency of the grating and the total instrument throughput can be very high.
Also this approach typically leads to relatively simple optical schemes with manufacturable optical components. However, the  increasing of spectral resolving power requires the increasing of grating frequency and the narrowing of spectral working range. In echelle-spectrographs built according to the second approach, the diffraction orders are separated by a cross-dispersion grating or prism. Thus, the spectral image consists of many lines
covering a wide range of wavelengths with high spectral resolving power. However, in addition to their technological complexity, such spectrographs are usually characterized by a lack of throughput.

For several research areas in modern astrophysics, it would be desirable to have an instrument occupying an intermediate position between the two above-mentioned groups in the resolution-throughput space. The authors had considered a possibility of building such a spectrograph on the basis of a cascade of volume-phase holographic (VPH) gratings \citep{MPV}. As explained below, such a spectrograph combines the high efficiency of VPHs with an improved spectral resolution over an extended spectral range, which becomes possible due to the registration of multiple spectral images at the same detector simultaneously. This concept is new for astronomical instrumentation, so it should be thoroughly investigated before implementing a full-scale instrument mounted on a real telescope. For this reason an optical scheme of a reduced and simplified laboratory prototype of the spectrograph was developed.

In this paper we consider an experimental study of the prototype. We should emphasize here that there are a number of uncertain factors (mainly those related to the holographic materials and recording) influencing the actual performance of the instrument, which must be tested experimentally.
The paper is organized as follows. Section 2 provides a short overview of the spectrograph optical design. Section 3 describes the manufacturing, alignment and experimental study of the prototype. The latter includes throughput tests and measurement of actual spectral resolution. We also provide a brief comparison of the prototype performance with that of existing instruments of different classes. Section 4 summarizes the lessons learned from building the prototype. In Section 5 scientific cases which can be addressed by such an instrument are presented.

\section{Concept of the instrument and optical design}

The concept of the instrument uses widely known properties of volume-phase holograms (VPHs). Such an optical element can provide high diffraction efficiency in a relatively narrow spectral region \citep{Caul}. Outside of this region most of the incident flux is transmitted to the 0$^{th}$ order of diffraction.
This feature can be used to couple several spectral and/or imaging channels in a single instrument as was shown, for instance, by \citet{MP}.
Particularly, if several VPH gratings are mounted one after another in a cascade, each of them can create a spectral image in a dedicated narrow spectral range without affecting the beam propagation in other wavelengths. In this case the spectral image will represent a corresponding number of lines covering an extended spectral range with high dispersion and spectral resolving power. With a proper choice of groove frequency and incidence angle for each grating, it is possible to obtain a constant linear dispersion in each line and achieve tangential line centering. The lines in a spectral image can be separated by rotating the gratings in the sagittal plane. The first version of such an optical scheme was proposed by \citet{MPV}, although a similar concept can be traced back to \citet{BAT}.

The spectrograph prototype considered in this paper is based on the optical scheme described and modeled in the studies of
\citet{MVF,MVFP}. For convenience, here we provide a recap of the scheme layout (see Fig.\,1) and its main parameters used previously for calculations and modeling. The spectrograph operates
in the visible range 430-680 nm. The operating region is divided into three bands: 430-513, 513-597 and 597-680 nm. The dispersion unit consists of three
VPH gratings operating in corresponding sub-ranges. The gratings spatial frequencies are 1726, 1523, and 1205 mm$^{-1}$, respectively. Below we refer to
them as Blue, Green, and Red gratings. Each of the gratings is imposed on a plane-parallel plate made of BK7 glass and protected by a cover glass (the
substrate thickness is 2.6 mm and the cover glass thickness is 2.2 mm). The length of each spectral image line is 20 mm (so the reciprocal linear dispersion
is 4.15 nm/mm). The limits on the line spacing in a spectral image are 1.5-4 mm. The gratings are consequently mounted in collimated beams (the Blue grating
is the first, and the Red grating is the last), so two identical commercial Tessar-type lenses are used as a collimator and a camera. Each of the lenses has a focal
length of 135 mm and an f-ratio of 2.8. The actual entrance f-ratio is assumed to be decreased to 4. It was shown previously that the scheme provides a
spectral resolution of 0.125-0.203 nm, 0.125-0.330 nm, and 0.125-0.151 nm in the three subranges specified above, respectively. The spectral resolving  power obtained by modeling reaches R=5124. The spectral resolution is
limited mainly by camera aberrations and varies significantly across the image plane. More details about the spectral resolution computation can be
found in \citet{MVF}.

\section{The prototype}
\label{Proto}

When creating the spectrograph prototype with limited available funds, we focused on studying the most important features of the instrument
-- its high throughput, spectral resolving power, and ability to simultaneously register a wide spectral range with a single standard 2K x 2K CCD.
The diameter of the collimated beam was chosen to be 40\,mm in order to decrease the cost of customized VPH-gratings.
(At the same time, such a diameter still allows us to reach the required spectral resolution with a 30$\mu m$ entrance slit).

The grating plates did not have an anti-reflective coating. This mitigation allowed us to reduce the total cost of the spectrograph and the time required
to build it. At the same time, reflection losses can be easily accounted for when analyzing the measured data. However, later we shall
return to the reflection losses question when discussing the full-scale instrument.

The VPH gratings were manufactured in the State Institute of Applied Optics (Kazan, Russia). Each of the gratings was recorded by two parallel
beams obtained from a single laser source on a layer of dichromated gelatin (DCG, see, e.g., \citet{NW}) and then protected by a cover glass.
The diffraction efficiency of such a grating depends on the thickness of holographic layer, the modulation of refractive index, and the slanting angle of fringes. Recording a customized grating requires precise control over all of these parameters at the same time.

Two identical Tessar-type commercial lenses were used as the camera and collimator. We used the widely available Tair-11 lenses (F=135 mm,
F/2.8). This lens has a proper focal length and an entrance pupil large enough to cover the dispersed beam. The lens aberrations are sufficiently small and
their exact values are known in advance. The collimator lens was diaphragmed down to F/4.

All these optical parts were assembled and adjusted in a duraluminium box coupled with a 2K x 2K, 13.5\,$\mu m$  pixel size CCD. Fig.\,2 shows the
grating assembly as integrated into the instrument. A fiber assembly coupled with a 30$\mu m$ slit and an aperture equal to the collimator
aperture was used in all our experiments presented below.

\subsection{Analysis of throughput}

An increased throughput is the key advantage of the proposed optical design. Consequently, its experimental measurement is the primary goal of our study.
The throughput was measured in a mounting which includes a stabilized calibration light source operating in a wide range of wavelengths, a standard
monochromator, and the spectrograph opto-mechanical unit and CCD. The throughput estimates were obtained by observing this light source at selected
wavelengths chosen by the monochromator in narrow (5-10 nm) bands with and without the spectrograph optics. We used the same CCD and equal exposure times to register the light in both cases. The resulting throughput values were obtained as fractions of the registered light energy after passing through
the optics, in relative units of the incoming light energy. The obtained results are presented in Fig.\,3. The lowest solid curve in Fig.\,3 illustrates the
measured throughput of the prototype.

As can be seen from the plot, the measured throughput is not so high compared to theoretical expectations \citep{MPV}. This can be explained within the
framework of our engineering simplifications. However, in the future, the use of high-quality collimating/projecting optics and effective anti-reflection
coatings will allow us to raise it above 50\% (the solid thick curve in Fig.\,3). Furthermore, this limitation is still not final due to
the fact that the efficiencies of the used gratings can be enhanced. The measured efficiencies of the gratings appeared to be lower than the values predicted
by computations. There are a few possible explanations. Firstly, as we mentioned before, the efficiency depends on the parameters of the holographic layer.
In our case these parameters could be made to have small errors during holographic recording or, more likely, they could be changed during the DCG post-processing or degraded with time. Secondly, the angles of incidence in the two planes define, on one hand, the resolution and mutual positions of
the image lines and, on the other hand, the grating efficiency curves. Therefore, during the instrument alignment the condition of maximum efficiency could be broken to keep the resolution. We also should note here that gratings are never tested under conical diffraction conditions at the manufacturing stage.
However, the found mismatch in the holographic layer parameters and/or replay angles is not a fundamental limitation of this design. In Section 4 we
consider possible ways to avoid it in the future.

We should note that the presented results include only the transmittance of the spectrograph optics, with no account for the light
loss due to atmospheric extinction, telescope reflectance, or light cutoff at the fiber input/output. The actual mounting conditions of the
future full-scale instrument are not defined yet, nor are its specifications and detailed optical design. This makes an assessment of the
on-sky performance of the instrument almost impossible and complicates comparison with the existing instruments. However, for the  better understanding of the proposed
spectrograph advantages, we compare the corrected experimental throughput with that of a number of existing instruments.
We took the throughput data sets for Keck-HIRES \citep{Vogt} and X-shooter at VLT \citep{Vernet} and corrected them for losses in the telescope optics \citep{Bass}. Similarly, we used the data for BTA-Scorpio corrected for the CCD quantum
efficiency \citep{AM}.  Finally, we corrected the data for Subaru FOCAS \citep{NE} for telescope losses and atmospheric extinction \citep{AE}.
The comparison
of the throughput curves is shown in Fig.\,4. HIRES demonstrate a performance, which is typical for echelle spectrographs. Its typical spectral resolving power is 67000, while the throughput is relatively low.
X-shooter is an unique spectrograph constructed by a consortium of Institutes from a number of countries. It covers the entire wavelength range from UV to IR and shows probably the best performance for this class of instruments. For the spectral range under consideration it provides the spectral resolving power of 3300-9100 in the UVB and 5400-17400 in the VIS, where the exact value depends on the slit width.
The typical value for Scorpio with VHPG550 is R=500. FOCAS provides R=2500 with the VPH650 grism and R=3000 with VPH520. From this plot one can see the niche between the existing types of spectral instruments, which corresponds to our prototype.
When compared to the classical high-resolution echelle spectrograph, our solution provides a substantially higher throughput,
but the spectral resolving power is lower (see the next subsection for the prototype spectral resolving power). Compared to a typical low-resolution spectrograph, it has a slightly lower throughput due a larger number of optical components and a complicated alignment process. These two factors will be present even if we exclude the imperfections of the gratings. However, it has a higher resolving power.
In comparison with the recent highly efficient echelle-based design used in X-shooter, the prototype shows similar maximum throughput and
less perturbations of the curve,  therefore the effective throughput is considerably higher. From the point of spectral resolution the
performance is approaching to that of the echelle spectrograph, though the exact ratio strongly depends on assumptions about sampling. We would like to
emphasize that this performance was achieved with relatively simple optical components and can be significantly improved in the future.
Finally, the throughput curves for our spectrograph are close to those of the FOCAS instrument, while the spectral resolving powers are comparable. Nevertheless, in contrast with FOCAS, the prototype allows us to register all the spectra simultaneously at the same detector. To achieve this capability, we must reconcile with the irregularity of the throughput. We have to use a narrow-band grating to minimize the gratings crosstalk. As a result, the throughput drops around the sub-range boundaries.
Relying on these comparative study we propose a few application strategies for our concept, which are described in Section 4.

\subsection{Analysis of spectral resolution and ghost levels}

Next, we obtained spectra of scattered solar light fed through the fiber assembly to the described spectrograph prototype. One such image is shown in Fig\,5. A vertical section of this image taken at the frame center (Fig.\,6) illustrates the relative intensities of the spectral bands accumulated with each of the gratings. The difference between the bands is explained by the grating manufacturing and alignment issues mentioned above. Moreover, the grating efficiencies are convolved with the CCD quantum efficiency and the light source intensity.

The cascaded design of the dispersive unit can suffer of secondary spectral images, which appear because of the presence of non-working diffraction
orders and overlapping efficiency curves of individual gratings. Hereafter we refer to these undesired images as 'ghosts', as adopted in spectroscopy,
even though their nature is different from that of the ordinary spectral ghosts.

In order to extract the observed solar spectra, we reduced the CCD frames in a standard manner including the following steps: cosmic-ray hit removal,
electronic bias subtraction, scattered light subtraction, spectrum extraction, etc. The wavelength calibration procedure was carried out using a list of
known solar spectral lines. The DECH software{\footnote{www.gazinur.com}} was used to process all the steps. As an output result we obtained a series of
solar spectra that we used to study the possible influence of ghosts from secondary reflections and to measure spectral resolution. An example of the output
solar spectra is shown in Fig.\,7. Comparing the standard high-resolution solar spectrum with the observed one, as well as analyzing closely located
spectral doublets, we estimate that the measured spectral resolving power is generally lower than the modeling results and theoretical predictions.
The best spectral resolving power of R=4000-4100 was obtained for the Blue
band at 500-510 nm. The corresponding theoretical value is 1905, whereas for the entire Blue band it varies between 1898 and 3776.
The highest value in the Green band is R=2600 at 544-555 nm. The theoretical value for this narrow region is 3944, and for the Green band
it is 1916-4460. The Red band exhibits R=3500 at 572-582 nm. The corresponding theoretical value is 3994 and for the whole band it is 3953-5124.
The actual image plane for each of the bands is tilted and shifted relative to the other bands, so alignment of the detector implies searching for the best-fit
plane. Thus, the difference between the observed and computed values for the Green and Blue gratings is due to the redistribution of the aberrations between
the green and blue image lines.  So the result for these two gratings can be considered acceptable. As to the observed results for the Red
grating we assume that the recording wavefronts could have some imperfections like a small defocusing and/or tilt, which affect the image quality and the
line centering. The suggestions on the dispersion unit modifications listed in Section 4 could help to eliminate such effects.
In order to provide a visual demonstration and quantitative estimation for the misalignments causing the observed resolution changes we
performed the following modelling. We optimized the values of angular positions of all the gratings, the dispersing unit positions, the lenses focusing and the Red grating residual optical power to fit the measured instrument functions widths. It was found that the observed spectral
resolution may occur with the following alignment errors: the collimator focusing error is 0.59 mm, the collimated beam decentering is 10.4 mm;
the dispersing unit angular position errors (X/Y) are 2.72/1.01$^\circ$; the Blue grating angular position errors are 0.08/-1.00$^\circ$,
the errors for the Green grating are -1.00/-1.01$^\circ$ and for the Red grating they are equal to 0.67/0.99$^\circ$; the Red grating residual
focus is 32.3 m, the camera focusing error is -0.69 mm, the CCD decentred by 0.98 mm and rotated by -1.03$^\circ$ around the X axis. Note that
these values are not actual measured ones, but they can provide a scale for the alignment problem. Presumably, these deviations are due to the search for a
compromise between mutually contradicting conditions of the maximum resolution and the maximum efficiency, as was explained before.

Visual inspection of the morphology of the scattered light out of the working spectral orders has revealed the presence of traces of spectral ghosts
at typical levels of considerably less than 0.5\% (zoomed area in Fig.\,6 demonstrates the strongest ghost). These ghosts are due to secondary reflections
and non-zero residual spectral orders other than the used first order (see explanation above). Meanwhile, such a level is almost perfect.
This is probably the most important result in our study which allows us to conclude that the use of a cascade of holographic gratings in astronomical
spectrographs enables one to obtain very high quality spectral data with signal-to-noise (S/N) ratios of up to S/N~=~1000.
We also note that this result should be considered as a lower limit due to the simplifications made in our prototype. We are sure that a
high-quality solution would allow us to obtain several-fold better results.

To summarize the analysis of experimental data given above we can conclude that the results obtained with the prototype should be considered as a successful proof of the concept. They completely confirm all key features of a spectrograph based on cascade of VPH gratings. Further design
enhancements and potential scientific applications are discussed below.

\section{Lessons learned and notes about a full-scale instrument}

The experience obtained during the development, construction, alignment and operation of the spectrograph prototype allows us to make a few conclusions
concerning the implementation of this or similar designs in the future:

\begin{itemize}
 \item
It may be reasonable to arrange the gratings in a red-to-blue order cascade. This may help decrease the channels crosstalk \citep{MVFPD},
although some geometrical conflicts might arise.

\item
Simultaneous focusing of several spectra onto the same detector can cause significant difficulties. At least some of them can be eliminated by
proper tolerance designation and wavefront control during the hologram recording.

\item
Reflection on multiple surfaces leads to a notable decrease in total throughput. Therefore, a more complex (and costly) anti-reflection coating should
be used. Alternatively, the design could be changed in order to decrease the number of refracting surfaces.
\end{itemize}

We should note here that some of the difficulties with the prototype are attributable to the used holographic material, i.e. dichromated gelatin (DCG).
The DCG processing consists of several steps, which complicates the precise control of the grating efficiency curve. Moreover, DCG has a high sensitivity to
environmental conditions, so each grating should be protected by a cover glass (even in this case there is a high risk of grating degradation).
Most of these issues can be resolved in a design using modern photopolymer holographic materials proposed by \citet{ZA,ZB}. More information
on photopolymer properties can be found in \citet{FKB} and other publications by these authors. Among them we should emphasize the experimentally proven
possibility of calibrating the refractive index modulation. This property allows one to produce customized VPHs with a high efficiency in the dedicated
region. In addition, this advanced concept can be applied not only when building a new instrument, but also when upgrading an existing one.

Summarizing all the conceptual advantages and shortcomings of the cascade-VPH spectrograph and its features, which we encountered during our
work on the prototype and this paper, we can propose the following strategies for building of a full-scale spectrograph:
\begin{itemize}
\item
\textit{Building of a simplified version of a multichannel VPH-spectrograph}. With the presented concept it is possible to create a high-performance instrument analogous to the leading VPH-based designs like \citet{NE} and \citet{NT}. The cascaded design allows to reach high throughput and moderately high spectral resolution over the entire working range at the same time. The spectrograph uses only one camera and a single detector and has no customized dichroic splitters. A spectrograph similar to the described prototype would be much cheaper and simpler, so it could be of special interest for a large number of observatories owning mid-class telescopes. However, our solution have some obvious limitations like the throughput irregularity shown above or possibility to implement some special observational modes like the integral field measurements.

\item
\textit{Competing with low-to-medium resolution echelle spectrographs}. As was demonstrated above, the proposed design has a visible difference from echelle spectrographs. An interested reader can easily estimate the difference in performance between our prototype and a potential
equivalent echelle spectrograph. Even the best available commercial echelle grating \citet{Nelson} has a limited efficiency, which doesn't exceed 58\,\%.
Using a typical transmission of a cross-disperser prism \citet{Hearnshaw} and assuming that the spectrograph has a simple collimating mirror and the camera
identical to the one used in our experiments one can find that for an equivalent echelle-based scheme the maximum throughput is 38.5\,\% at the best.
Meanwhile, with a blaze angle of 63$^\circ$ (\citet{Nelson}) and a focal ratio of 4 such a spectrograph could reach R7600 at the 6-m telescope.
Moreover, the using of a more complex optics or a further mitigation of the spectral resolution would not help to achieve a significant increase
of the echelle spectrograph throughput. On contrary,
the using of an advanced camera and AR coatings would allow to build a VPH-based instrument with throughput approaching the maximum of ~ 60\,\%
shown by modelling \citet{MVFP}.
In addition, different target specifications would require a customized (and therefore, much more expensive) echelle. A customized VPH can be produced
with exactly the same equipment and materials as a standard one, so the cost difference will be relatively small.
Thus, we can conclude that although for high-resolution application echelle spectrographs are out of competition, for the given niche application a
cascaded VPH has significant advantages. Below we show a number of science cases when an increased throughput is a critical requirement, while the spectral
resolving power of R~5000 is enough.

\item
\textit{Replacement of existing spectrographs with transmitting dispersers.} There is a large number of obsolete low-to-medium resolution spectrographs
using VPH gratings and grisms. They can be re-built with a new cascaded or multiplexed VPH disperser. In the majority of cases some parts of the
existing instrument, especially, the camera and collimator optics and the detector assembly, can be re-used. This strategy is of a special interest
for instruments, which already use interchangeable VPH dispersers like \citet{Clemens} or the above-mentioned \citet{AM}. In this case one can obtain
actually a new instrument with a substantial gain in performance. For instance, thanks to the simultaneous registration of spectra one can obtain a
substantial gain in the observation time. On the other hand, the expenses in terms of design and manufacturing would be minimal. We must note that a more
advanced version of this strategy was already proposed and explored in \citet{ZA,ZB}. In that case even the geometry and all the mechanical parts of the
instrument can be kept.
\end{itemize}

\section{Science with the spectrograph}

The presented spectrograph has a number of advantages in comparison with existing spectrographs between low-resolution instruments
(the resolving power less or about R1000 for the whole visible range, a total throughput higher than 50\,\%) and high-resolution
echelle spectrographs (the resolving power higher than R10 000, a total throughput less than or about 10\,\%). Our cascade-VPH
scheme has advantages in both above-mentioned types of spectrographs, it can achieve a resolving power of at least R5000 and a total
throughput up to 50 \%. This solution provides comparatively high resolving power and keeps a wide spectral region typical of echelle-spectroscopy
 with much higher throughput. Naturally, the  resolving power of the prototype is lower than that of echelle-spectrographs.

This spectrograph is an intermediate solution between low-resolution spectrographs with single grism/VPH  and  echelle spectrographs.
In the first case one can observe very faint targets with a resolving power of R1000 - R2000 (150 - 300 km/sec). In the second case, by using
echelle-based spectrographs we can not detect fainter targets. The echelle spectrograph covering the whole optical range can detect stars only to
15th magnitude, or up to the 18th magnitude if we use the telescopes of the class 8-10m. Therefore, a cascade-VPH spectrograph is more effective
in the studies of faint and low-contrast objects.

The main advantage of the cascade-VPH prototype is a possibility to study low-contrast and faint objects: i) a higher resolution power of about 60
km/sec, ii) a total throughput of up to 50 \%, and iii) the wide spectral range.\\

{\it Requirement on the spectral resolution.}\\

In such a case we may really distinguish H\,II regions and planetary nebulae from stellar winds in massive stars. Massive stars on the Main Sequence and
beyond, at the final stages of their evolution, are observed in galaxies in the Local Universe at distances of up to 50 Mpc; as a rule they are observed
in a strong background (H\,II regions). They are LBV stars (Luminous Blue Variables), B[e]-supergiants, WR-stars
\citep{Humphrey1994,Fabrika2005,Sholukhova2011,Neugent2012,Sholukhova2015}; their magnitudes are fainter than 18. LBV stars in their cool stage may
have emission line widths (FWHM) of about 150 - 200 km/sec. They produce the surrounding nebulae as a result of mass loss due to strong winds. It is
necessary to distinguish between the nebula and the winds. The same is valid for a search of objects in nearby galaxies
\citep{Fabrika1995}, it is an important clue for understanding these targets. X-ray sources in galaxies also produce jets and nebulae; the large nebula
S26 in the nearby galaxy NGC\,7793 \citep{Pakull2010}, which looks like a copy of the famous Galactic source SS\,433
\citep{Fabrika1997}, the only known super-Eddington accretion disk in the Galaxy, is one example.
In our Galaxy the black hole binaries (accretion disks) have the dispersion velocity \citep{Soria1999,Rahoui2017}
from 400 to 1000 km/sec; in the case of surrounding nebulae, a cascade-VPH spectrograph can reveal the difference between broad
emission lines ans H\,II regions. {\bf We note that besides spectral resolution other items (a throughput and spectral ranges are also very
important).}

For bright stars in the Galaxy the main advantage is a good spectral resolution and a highest throughput.
Magnetic field up to 100 Gauss may be measured for brighter stars \citep{Monin2002} with a higher S/N ratio.
If we look at spectra of standard DA white dwarfs, one can not distinguish narrow absorption lines appearing in the centers of the hydrogen lines
(H$\beta$, H$\alpha$), but using the whole range with the cascade-VPH spectrograph one can implement such a resolution. The same is valid for
magnetic white dwarfs using the polarimetry. It is also necessary to cover the whole visible range, where hydrogen or helium lines may shift to
hundreds of angstroms with unexpected different broadenings. Especially interesting is the fast line-profile variability in the spectra
of OB stars \citep{Kholtygin2003}; these stars are bright, and in a few minutes of exposure the S/N ratio can reach about 1000. Therefore one needs
to have both a good transmission and a good spectral resolution for these stars.\\

{\it Requirement on the throughput.}\\

Recently, objects of a new type --- ultraluminous X-ray sources (ULXs) --- have been discovered in nearby galaxies. {\bf It is important to have a
good throughput and as well the resolution and the optical range. These are super-Eddington accretion disks or intermediate mass black holes (IMBHs). Optical
spectroscopy of these targets (21-24th magnitudes) is very significant} \citep{Cseh2013,Fabrika2015}, as there are only a few such spectra, since
these objects are very faint. Again, they are located against a strong background of H\,II regions with different ionizations and different
velocities. It is possible to distinguish between the H\,II regions ($ O[III] \lambda \lambda 4959, 5007$, N[II] $\lambda \lambda 6548, 6583$,
S[II] $\lambda \lambda 6716, 6730$) and hydrogen, He\,I and He\,II emission lines in ULXs. In star-forming galaxies, in centers of young stellar
clusters very massive stars (VMSs) with masses higher than 200 solar masses \citep{Crowther2010,Solovyeva2017} can produce IMBHs in a short evolution
time. It is necessary to have the whole spectral range with a good resolution, because in the blue and yellow ranges we observe high-ionization nebular
lines ([OIII]), in the red region the low-ionization lines ([OI], [NII], [SII]), they may be different in their structure and velocities.
In the same young stellar clusters it is difficult to find high- and low-ionization nebulae with different dispersion velocities. It could
be critically important, because one can detect both narrow and broad lines, as in emission and in absorption. This is a new field in the modern
astrophysics.

Finally, a cascade-VPH spectrograph can be effectively used in the studies of exoplanets. In particular, it will be a very effective instrument
in a hunt for scattered light from hot-jupiter exoplanets \citep{G1}, and for spectral studies of transit events caused by giant planets
\citep{Lang,G}. These studies are presently among the hottest topics of the modern astrophysics.\\

{\it Requirement on the spectral range.}\\

In nearby galaxies like M\,33, M\,31, M\,81 there are many {\bf objects like novae}. Besides massive stars and X-ray sources one observes novae and recurrent
novae \citep{Darnley2016}, which can appear as bright targets of 16-17 magnitude and then weaken in a week or more. The observations of
novae show that their emission lines become narrower with time, dropping from several thousands km/sec to 100 km/s (depending on the  ionization state) as their
brightness decreases.

When we go to our Galaxy with its brighter targets up to 18th magnitude, the main reasons are the spectral resolution up to R5000
and the wide spectral range $\sim 450 - 680$\,nm. The presented spectrograph can also be effectively used in
spectrophotometric studies of photometrically variable magnetic white dwarfs \citep{Brink,Val1,Val2}, as well as in their
moderate-resolution spectropolarimetric studies \citep{Valyavin2006,L1} by equipping the spectrograph with polarimetric optics \citep{VAN}.
In magnetic white dwarfs, a search for kilogauss magnetic fields in white dwarfs and hot subdwarf stars can be performed. Using spectropolarimetry
with such a resolution allows one to observe stars like T Tauri stars \citep{Smirnov2004} when both emission and absorption lines can be
measured.

The spectrograph can be easily updated to use different spectral ranges with the same spectral resolution and throughput. The main spectral
features must be shifted to have maximum (or nearly maximum) throughput. They are H$\gamma$, H$\beta$, H$\alpha$, the main He\,I lines ($\lambda \lambda
4471, 4922, 5015, 5876, 6678$), the Bowen blend CIII/NIII ($\sim \lambda 4640$), the He\,II line $\lambda 4685$, and the main nebular lines $ O[III] \lambda
\lambda 4959, 5007$, N[II] $\lambda \lambda 6548, 6583$, S[II] $\lambda \lambda 6716, 6730$). It is also possible to install an additional VPH in
the cascade to make a broader spectral range.

\section{ACKNOWLEDGEMENTS}

\begin{acknowledgements}
The creation and characterization of the prototype were supported by the Russian Science Foundation (grant N14-50-00043).
S.F. acknowledges the RFBR grant N16-02-00567 for support of the science with the spectrograph, and the Russian Government Program of Competitive Growth of KFU. E.M. acknowledges the support of his personal post-doc contract from the European Research council through the H2020 - ERCSTG-2015 - 678777 ICARUS program.
\end{acknowledgements}

\newpage

\begin{figure}
    \centering
    \begin{center}
 \includegraphics[width=6.85cm]{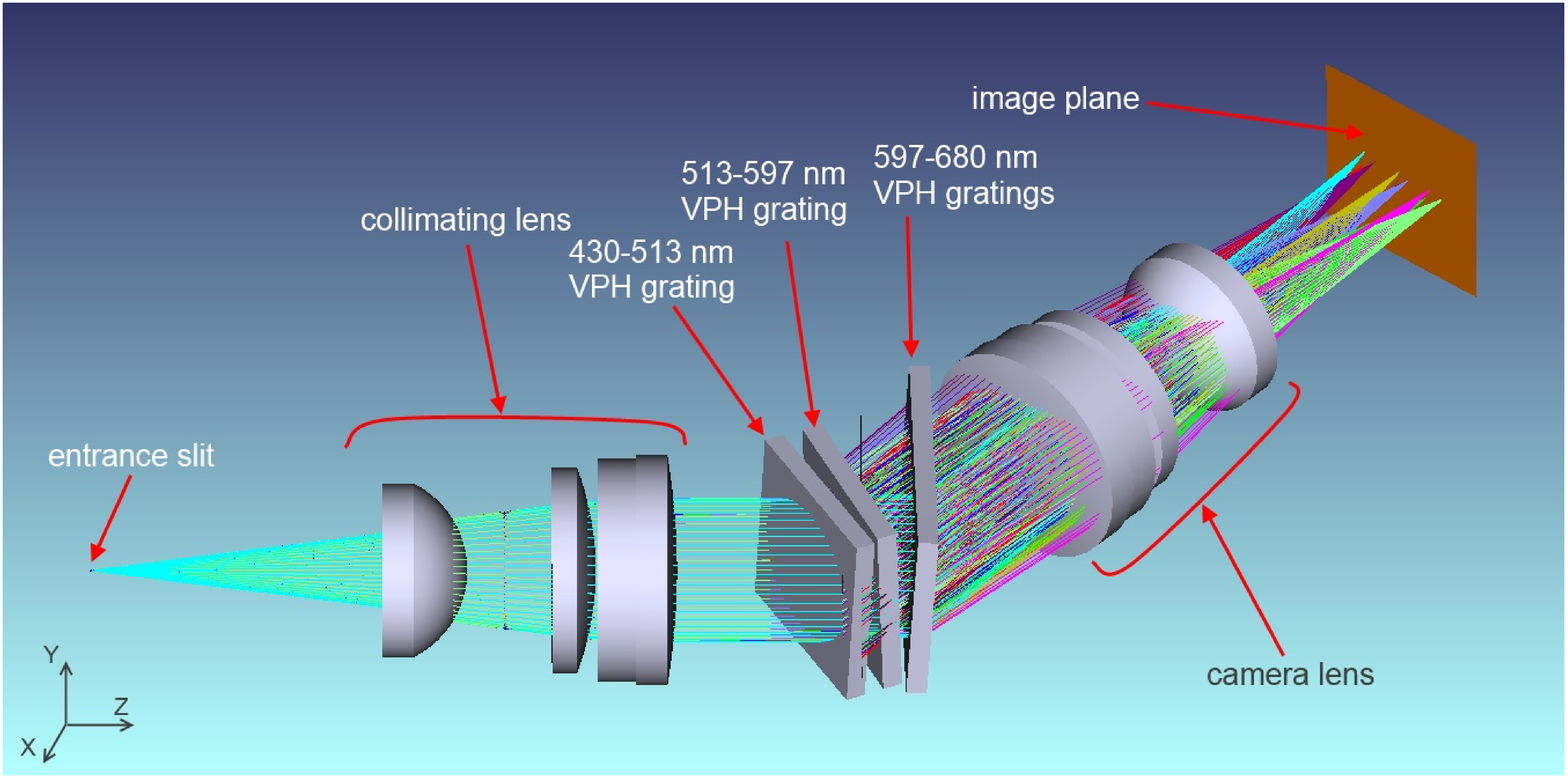}
\end{center}
    \caption{General view of the spectrograph prototype optical scheme.}
\label{fig1}
\end{figure}

\begin{figure}
    \centering
    \begin{center}
 \includegraphics[width=5.85cm]{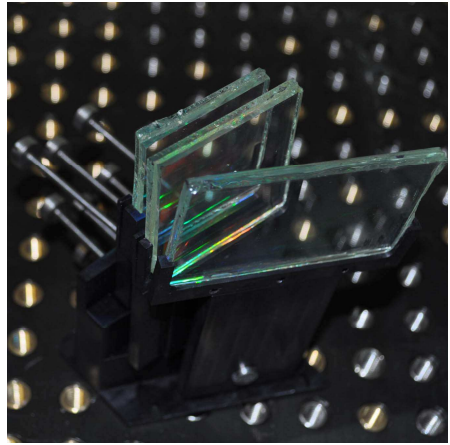}
\end{center}
    \caption{Assembled gratings. From left to right: blue, green and
    red.}
\label{fig2}
\end{figure}

\begin{figure}
    \centering
    \begin{center}
\hspace*{-2.5cm}
\includegraphics[width=12cm]{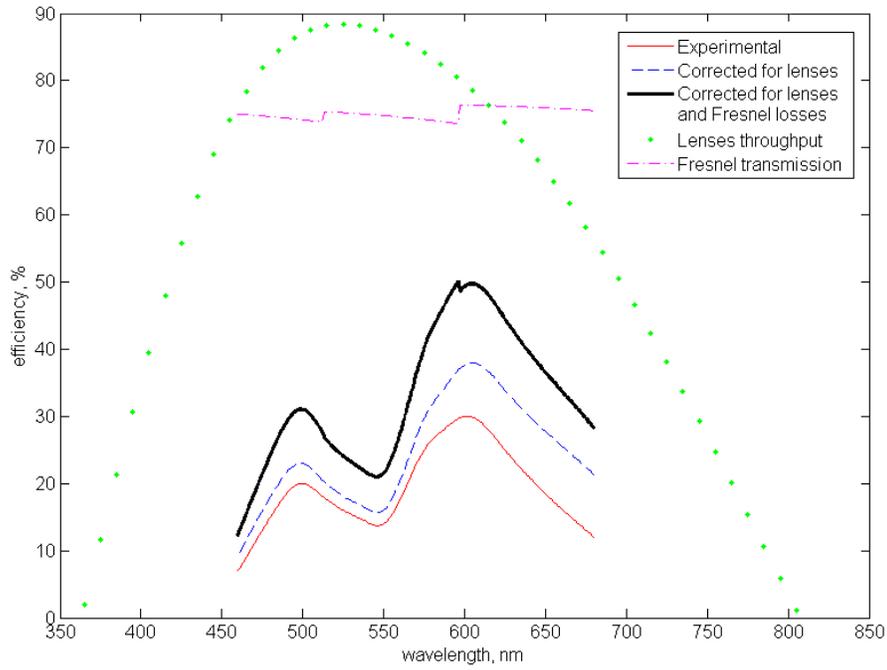}
\end{center}
    \caption{Efficiency of individual optical parts and reconstructed total throughput of the spectrograph (black solid curve).}
\label{fig3}
\end{figure}

\begin{figure}
    \centering
    \begin{center}
\hspace*{-2.5cm}
\includegraphics[width=20cm]{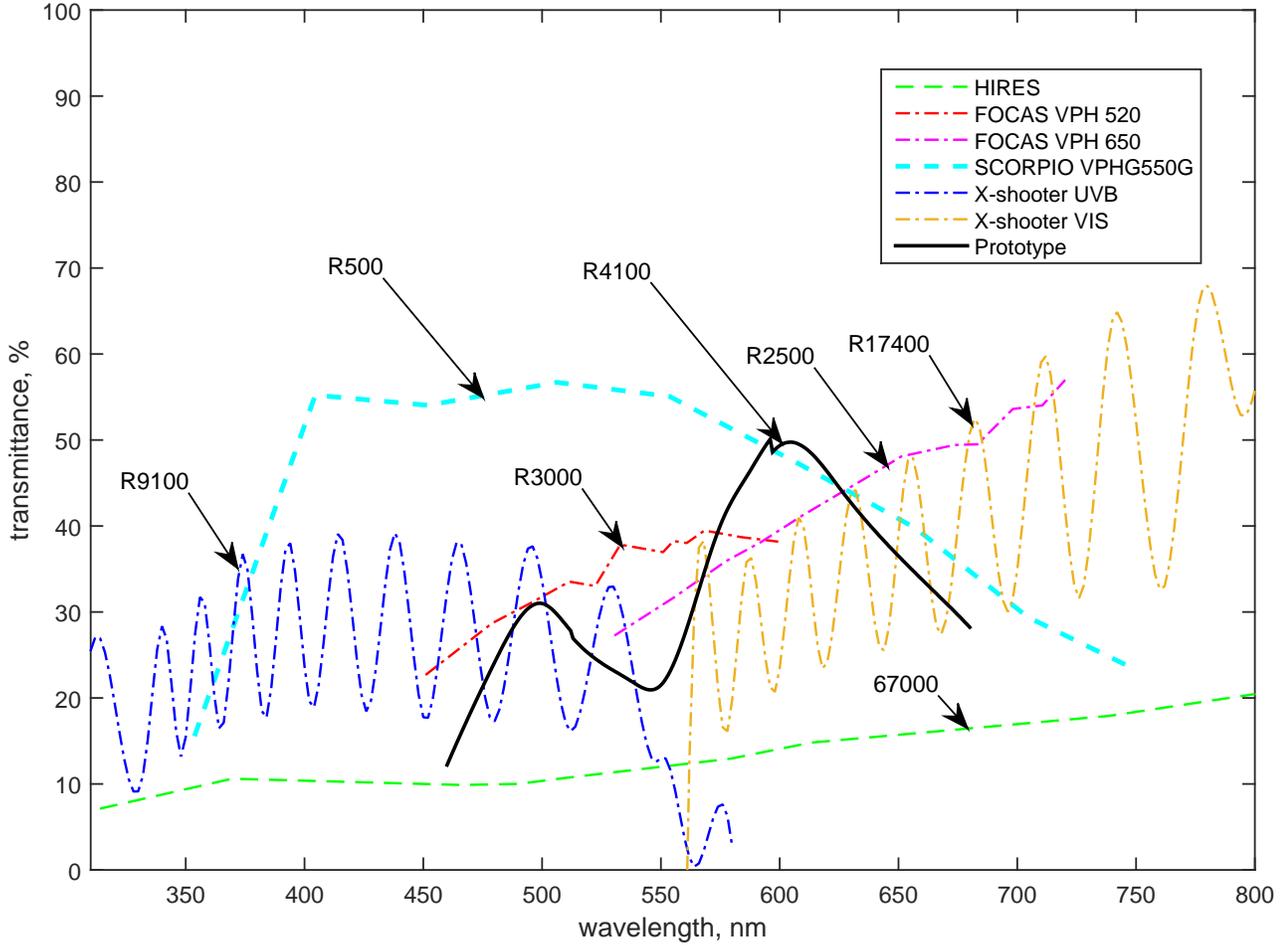}
\end{center}
    \caption{\textbf{Comparison of the corrected throughput of the prototype with those of existing astronomical spectrographs of different classes.}}
\label{fig4}
\end{figure}

\begin{figure}
    \centering
    \begin{center}
 \includegraphics[width=8cm, angle=270]{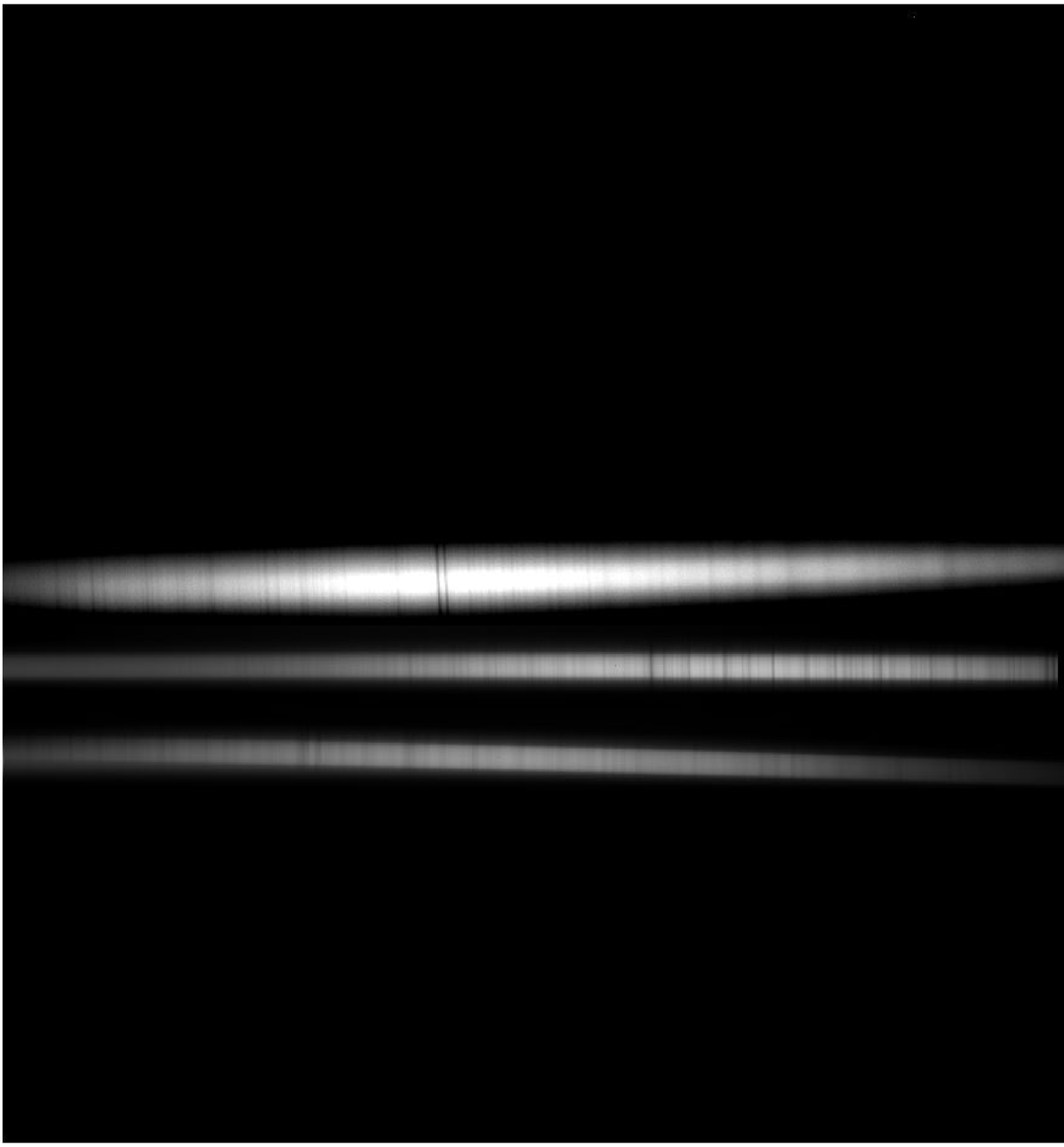}
\end{center}
    \caption{CCD-image of the solar spectrum.
From bottom to top: Blue, Green, and Red.
}
\label{fig5}
\end{figure}

\begin{figure}
    \centering
    \begin{center}
 \includegraphics[width=15cm]{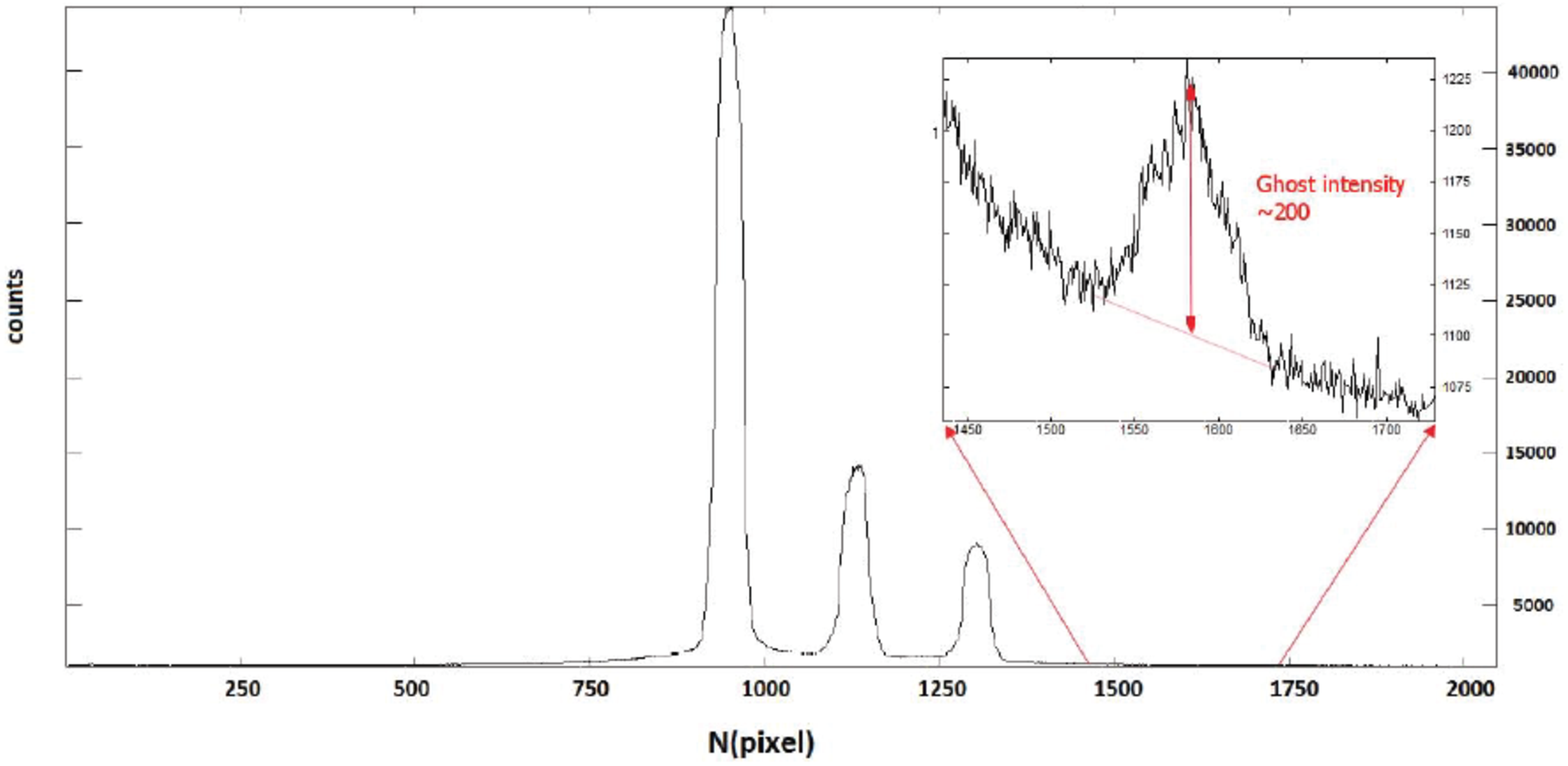}
\end{center}
    \caption{Vertical section of the CCD-image of the solar spectrum}
\label{fig6}
\end{figure}

\newpage

\begin{figure}
    \centering
    \begin{center}
    \includegraphics[width=15cm]{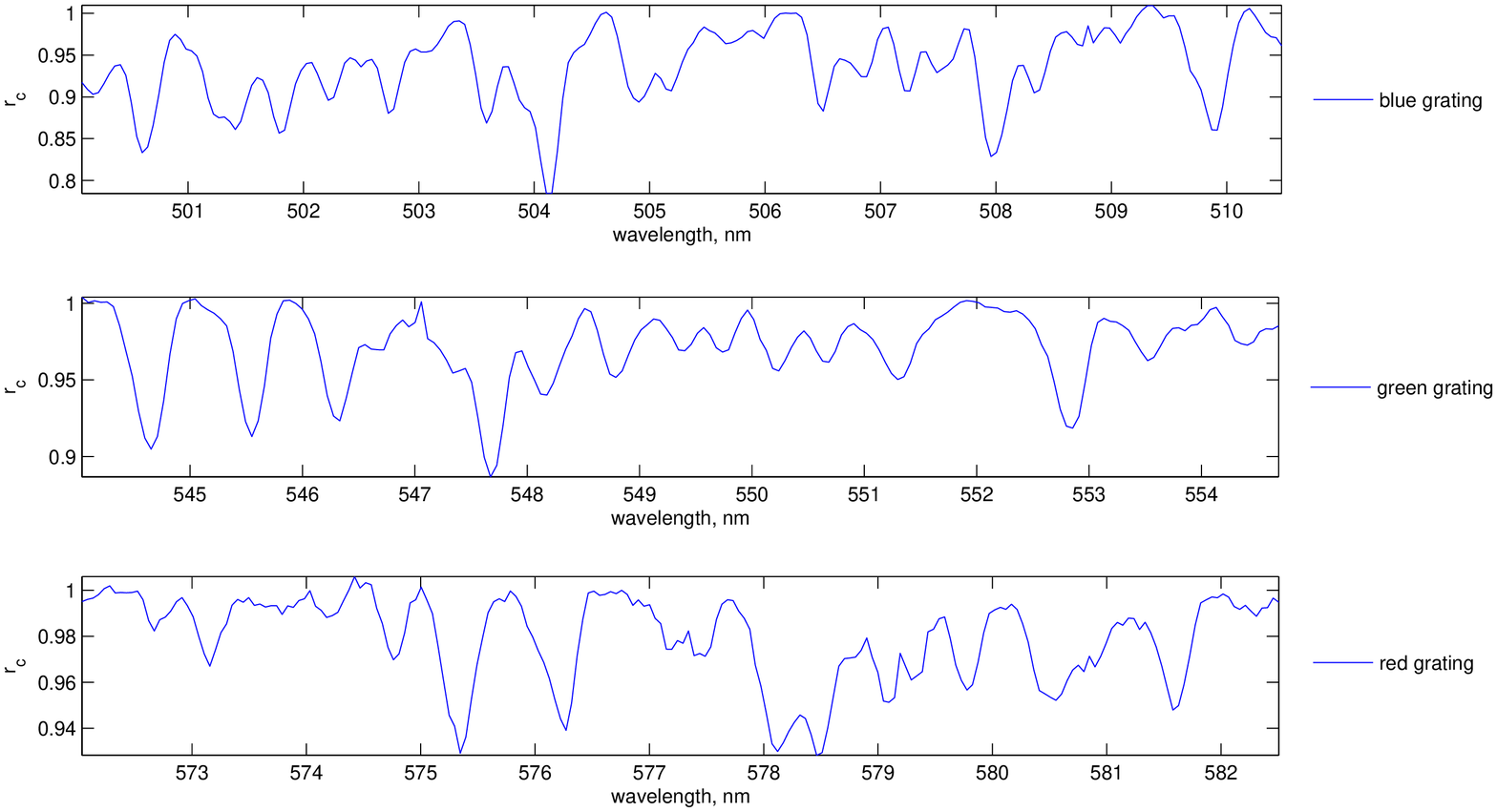}
\end{center}
    \caption{Fragments of normalized wavelength-calibrated solar spectra obtained
    with the spectrograph prototype for all three gratings.}
\label{fig7}
\end{figure}


\begin{thebibliography}{99}

\bibitem[Afanasiev and Moiseev (2005)]{AM}
Afanasiev, V. L., Moiseev, A. V. 2005, Astronomy Letters, 31, 194

\bibitem[Bass (2014)]{Bass}
Bass, M. (ed.) 1995, Handbook of Optics ,McGraw-Hill, New York

\bibitem[Battey et al. (1995)]{BAT}
Battey, D.E., Owen, H., Tedesco, J.M. 1995, US Patent No.5, 442, 439.

\bibitem[Brinkworth et al. (2013)]{Brink}
Brinkworth, Carolyn S., Burleigh, Matthew R., Lawrie, Katherine, Marsh, Thomas R., and
Knigge, Christian 2013, ApJ, 773,  article id. 47, 16 pp.

\bibitem[Bruder et al. (2010)]{FKB}
Bruder Friedrich-Karl, Deuber Francois, Fäcke Thomas, Hagen Rainer et al. 2010,
Proc. SPIE, Practical Holography XXIV: Materials and Applications, 7619, 76190I.

\bibitem[Buton et al. (2013)]{AE}
Buton, C., Copin, Y., Aldering, G., Antilogus, P. et al. 2013, A\&A, 549, A8

\bibitem[Caulfield (1979)]{Caul}
Caulfield, H.J. 1979, Handbook of optical holography, Academic Press, New York

\bibitem[Clemens (2004)]{Clemens}
Clemens, J. C., Crain, J. A., Anderson R. 2004, Proc. SPIE 5492,
Ground-based Instrumentation for Astronomy

\bibitem[Crowther et al. (2010)]{Crowther2010}
Crowther, P. A., Schnurr, O., Hirschi, R., Yusof, N., Parker, R. J. Goodwin, S. P. and Kassim, H. A. 2010, MNRAS, 408, 731

\bibitem[Cseh et al. (2013)]{Cseh2013}
Cseh, D., Grisé, F., Kaaret, P., Corbel, S., Scaringi, S., Groot, P., Falcke, H. and Körding, E. 2013, MNRAS, 435, 2896

\bibitem[Darnley et al. (2016)]{Darnley2016}
Darnley, M. J. et al. 2016, ApJ, 833, article id. 149, 38 pp.

\bibitem[Ebizuka et al. (2011)]{NE}
Ebizuka Noboru, Kawabata Koji S., Oka Keiko, Yamada Akiko et al., 2011,
Publications of the Astronomical Society of Japan, 63, S613

\bibitem[Fabrika (1997)]{Fabrika1997}
Fabrika, S. N. 1997, Astrophys. and Space Sci., 252, 439

\bibitem[Fabrika and Sholukhova (1995)]{Fabrika1995}
Fabrika, S. and Sholukhova, O. 1995, Astrophys. and Space Sci., 226, 229

\bibitem[Fabrika (2005)]{Fabrika2005}
Fabrika, S., Sholukhova, O., Becker, T., Afanasiev, V., Roth, M. and Sanchez, S. F. 2005, A\&A, 437, 217

\bibitem[Fabrika et al. (2015)]{Fabrika2015}
Fabrika, S., Ueda, Y., Vinokurov, A., Sholukhova, O. and Shidatsu, M. 2015, Nature Physics, 11, 551

\bibitem[Grauzhanina et al. (2017)]{G}
Grauzhanina, A. O., Valyavin, G. G., Gadelshin, D. R., Baklanova, D. N., Plachinda, S. I., Antonyuk, K. A., Pit, N. V., Galazutdinov, G. A.,
Valeev, A. F., Burlakova, T. E., Kholtygin, A. F. 2017, Astrophysical Bulletin, 72, 67

\bibitem[Grauzhanina et al. (2015)]{G1}
Grauzhanina, A., Valyavin, G., Gadelshin, D., Zhuchkov, R., Galazutdinov, G., Burlakova, T., and  Mkrtichian, D.
2015, ASPC, Physics and Evolution of Magnetic and Related Stars, 494, 289

\bibitem[Hearnshaw (2009)]{Hearnshaw}
Hearnshaw J. E. 2009, Astronomical Spectrographs and their History, Cambridge University Press, Cambridge, UK

\bibitem[Humphreys and Davidson (1994)]{Humphrey1994}
Humphreys, R.M. and Davidson, K. 1994, PASP, 106, 1025

\bibitem[Kholtygin et al. (2003)]{Kholtygin2003}
Kholtygin, A. F., Monin, D. N., Surkov, A. E. and Fabrika, S. N. 2003, Astronomy Letters, 29, 175

\bibitem[Langford et al. (2011)]{Lang}
Langford, Sally V., Wyithe, J. Stuart B., Turner, Edwin L., Jenkins, Edward B., Narita, Norio, Liu, Xin, Suto, Yasushi, Yamada, Toru
2011, MNRAS, 415, 673

\bibitem[Landstreet et al. (2012)]{L1}
Landstreet, J. D., Bagnulo, S., Valyavin, G. G., Fossati, L., Jordan, S., Monin, D., and Wade, G. A.
2012, A\&A, 545A, 30L

\bibitem[Monin et al. (2002)]{Monin2002}
Monin, D. N., Fabrika, S. N. and Valyavin, G. G. 2002, A\&A, 396, 131

\bibitem[Muslimov et al. (2017)]{MVFP}
Muslimov, E., Valyavin, G., Fabrika, S., and Pavlycheva, N.  2017, Appl. Opt. 56, 4284

\bibitem[Muslimov et al. (2017)]{MVFPD}
Muslimov, E., Valyavin, G., Fabrika, S. and Pavlycheva, N. 2017, Proc. SPIE 10233, Holography: Advances and Modern Trends V, 102331M

\bibitem[Muslimov et al. (2016)]{MPV}
Muslimov, E., Pavlycheva, N.K., Valyavin, G.G., \& Fabrika, S.N. 2016, Astrophys. Bull. 71, 357.

\bibitem[Muslimov et al. (2016)]{MVF}
Muslimov, E., Valyavin, G.G., Fabrika, S.N., $\&$ Pavlycheva, N.K. 2016, Proc. SPIE 9908, Ground-based and Airborne Instrumentation for Astronomy VI,
990842.

\bibitem[Muslimov \& Pavlycheva (2015)]{MP}
Muslimov, E.R. \& Pavlycheva, N.K. 2015, Journal of the European Optical Society - Rapid publications, Europe, v. 10

\bibitem[Tamura et al. (2016)]{NT}
Tamura Naoyuki, Takato Naruhisa, Shimono Atsushi, Moritani Yuki et al.
2016, Proc. SPIE 9908, Ground-based and Airborne Instrumentation for Astronomy VI, 99081M

\bibitem[Nelson et al. (2010)]{Nelson}
Nelson Peter G., Casini Roberto, de Wijn Alfred G., Knoelker Michael 2010,
"The Visible Spectro-Polarimeter (ViSP) for the Advanced Technology
Solar Telescope," Proc. SPIE 7735, Ground-based and Airborne
Instrumentation for Astronomy III, 77358C (21 July 2010)

\bibitem[Naidenov et al. (2002)]{VAN}
Naidenov, I. D., Valyavin, G. G., Fabrika, S. N., Borisov, N. V., Burenkov, A. N., VikuliEv, N. A., Moiseev, S. V.,
Kudryavtsev, D. O., and Bychkov, V. D. 2002, 	Bulletin of the Special Astrophysical Observatory, 53, 124

\bibitem[Neugent et al. (2012)]{Neugent2012}
Neugent, K.F., Massey, P. and Georgy, C. 2012, ApJ, 759, 11

\bibitem[Newell (1987)]{NW}
Newell John Christopher William  1987, OPTICAL HOLOGRAPHY IN DICHROMATED GELATIN.,
St. Cross College, Oxford. A Thesis submitted for the Degree of Doctor of Philosophy at the
University of Oxford. Hilary Term.

\bibitem[Pakull et al. (2010)]{Pakull2010}
Pakull, M. W., Soria, R. and Motch, C. 2010, Nature, 466, 209

\bibitem[Palmer (2014)]{Palm}
Palmer, C. and Loewen, E. 2014, Diffraction gratings handbook, Newport Corporation, Rochester

\bibitem[Rahoui et al. (2017)] {Rahoui2017}
Rahoui, F., Tomsick, J. A., Gandhi, P., Casella, P., Fürst, F., Natalucci, L., Rossi, A., Shaw, A. W., Testa, V., Walton, D. J.
2017, Monthly Notices of the Royal Astronomical Society, 465, 4468

\bibitem[Sholukhova et al. (2011)]{Sholukhova2011}
Sholukhova, O. N., Fabrika, S. N., Zharova, A. V., Valeev, A. F., Goranskij, V. P. 2011, Astrophysical Bulletin, 66, 123

\bibitem[Sholukhova et al. (2015)]{Sholukhova2015}
Sholukhova, O., Bizyaev, D., Fabrika, S., Sarkisyan, A., Malanushenko, V. and Valeev, A. 2015, MNRAS, 447, 2459

\bibitem[Smirnov et al. (2004)]{Smirnov2004}
Smirnov, D. A., Lamzin, S. A., Fabrika, S. N. and Chuntonov, G. A. 2004. Astronomy Letters, 30, 456

\bibitem[Solovyeva et al. (2017)]{Solovyeva2017}
Solovyeva, Y. N., Fabrika, S. N., Vinokurov, A. S., Sholukhova, O. N., Valeev, A. F. 2017, ASPC, Stars: From Collapse to Collapse, 510, 58

\bibitem[Soria et al. (1999)]{Soria1999}
Soria, R., Wu, K., Johnston, H. M. 1999, Monthly Notices of the Royal Astronomical Society, 310, 71

\bibitem[Valeev et al. (2015) ]{Val1}
Valeev, A. F., Antonyuk, K. A., Pit, N. V., Solovyev, V. Ya. et al. 2015, Astrophysical Bulletin, 70, 318

\bibitem[Valeev et al. (2017) ]{Val2}
Valeev, A. F., Antonyuk, K. A., Pit, N. V., Moskvitin S.A. et al. 2017, Astrophysical Bulletin, 72, 44

\bibitem[Valyavin et al. (2006)]{Valyavin2006}
Valyavin, G., Bagnulo, S., Fabrika, S.; Reisenegger, A., Wade, G. A., Han, Inwoo and Monin, D. 2006, ApJ, 648, 559

\bibitem[Vernet et al. (2011)]{Vernet}
Vernet, J.; Dekker, H.; D`Odorico, S.; Kaper, L. et al. 2011, Astronomy \& Astrophysics, 536, A105

\bibitem[Vogt et al. (1994)]{Vogt}
Vogt, S., Allen, S., Bigelow, B., Bresee L., et al. 1994,  Proc. SPIE 2198, 362

\bibitem[Zanutta et al. (2017a)]{ZA}
Zanutta, A., Orselli, E., Fäcke, T., and Bianco, A. 2017, Proc. SPIE 10233, Holography: Advances and Modern Trends V, 1023316

\bibitem[Zanutta et al. (2017b)]{ZB}
Zanutta, A. Landoni, M., Riva, M., and Bianco, A. 2017, MNRAS\textit{,  469, 2412 }

\end{thebibliography}
\end{document}